# Thermal Comptonization in Mildly Relativistic Pair Plasmas

J. G. Skibo[1] and C. D. Dermer
E. O. Hulburt Center for Space Research, Code 7653, Naval Research Laboratory, Washington, DC 20375-5352

R. Ramaty
Laboratory for High Energy Astrophysics, Goddard Spaceflight Center, Greenbelt, MD 20771

and

J. M. McKinley
4010 Granada Drive, S.E., Huntsville, AL 35802

## ABSTRACT

We use a Monte Carlo simulation to calculate the spectra of mildly relativistic thermal plasmas in pair balance. We use the exact integral expression for the electron-positron thermal annihilation spectrum, and provide accurate expressions for the Gaunt factors of electron-ion, electron-electron, and electron-positron thermal bremsstrahlung in the transrelativistic temperature regime. The particles are assumed to be uniformly distributed throughout a sphere, and the pair opacity is self-consistently calculated from the energy and angular distribution of scattered photons. The resultant photon spectra are compared with the nonrelativistic diffusion treatment of Sunyaev and Titarchuk, the bridging formulas of Zdziarski, and the relativistic corrections proposed by Titarchuk. We give a corrected formula for the spectral index resulting from Comptonization in the low temperature, low optical depth regime, and show where the effects of bremsstrahlung production are important in spectral calculations.

We calculate allowed pair-balanced states of thermal plasmas with no pair escape which include bremsstrahlung and internal soft photons. The results are presented in the spectral index/compactness plane, and can be directly compared with observations of spectra from AGNs and Galactic black hole candidates. By comparing with X-ray spectral indices of Seyfert AGNs and compactnesses inferred from X-ray variability data, we find that the allowed solutions for pair equilibrium plasma imply that the temperatures of Seyfert galaxies are $\lesssim$ 300 keV. This prediction can be tested with more sensitive gamma-ray observations of Seyfert galaxies. We find that if the X-ray variability time scale gives an accurate measure of the compactness, pair-dominated solutions are inconsistent with the data.

---

[1]NAS/NRC Research Associate

## 1. Introduction

Comptonization in mildly relativistic plasmas has long been thought to play an important role in producing the continuum emission from compact Galactic X-ray sources (Zel'dovich & Shakura 1969; Illarionov & Sunyaev 1972; Felten & Rees 1972). It has also been suspected that such plasmas reside near the innermost regions of Galactic black holes such as Cygnus X-1 (Shapiro, Lightman & Eardley 1976; Sunyaev & Titarchuk 1980). In addition to Comptonization, pair production and annihilation may also be important in such plasmas. In fact, the emission around 1 MeV reported from the direction of the Galactic center (Riegler et al. 1985) and Cygnus X-1 (Ling et al. 1987) was interpreted as radiation from pair-dominated plasmas (McKinley 1986; Liang & Dermer 1987). Furthermore, observations of transient features with energies near 0.4 MeV observed from Galactic black hole candidates (e.g. Gilfanov et al. 1994) could be the signature of pair annihilation, although other interpretations are possible (Skibo, Dermer, & Ramaty 1994).

Evidence for the existence of mildly relativistic thermal plasmas in extragalactic astrophysical environments is provided by X-ray and gamma-ray observations of the Seyfert class of Active Galactic Nuclei (AGN). Observations at X-ray energies made with HEAO-1 and EXOSAT revealed a power law spectral shape with a canonical photon spectral index $\Gamma \sim 1.7$ in the 2-10 keV range (Rothschild et al. 1983; Turner & Pounds 1989). More recent analyses of Ginga observations of Seyfert 1 galaxies, taking into account the effects of reflection by a cold optically-thick medium, imply that the mean X-ray spectral index is closer to $\Gamma \sim 1.9 - 2.0$ (Pounds et al. 1990; Nandra & Pounds 1994). At gamma-ray energies, observations of Seyfert galaxies made with OSSE on the Compton Gamma-Ray Observatory (CGRO) reveal spectra which generally display exponential cut offs or spectral softenings with an e-folding energy of $\sim$ 50 keV - several hundred keV (Maisack et al. 1993; Johnson et al. 1994; Zdziarski et al. 1995). These observations, in conjunction with the rapid temporal X-ray variability observed from these sources (Barr & Mushotzky 1986; McHardy 1989; Done & Fabian 1989; Grandi et al. 1992), strongly suggest the presence of mildly relativistic thermal plasmas in the cores of Seyfert galaxies.

For pair-free plasmas at nonrelativistic temperatures and high optical depths, analytic formulae for the Comptonized spectra were derived by Sunyaev and Titarchuk (1980, hereafter ST80) assuming the presence of a low-energy photon source with luminosity greatly exceeding the intrinsic bremsstrahlung luminosity of the plasma. This problem was treated at relativistic temperatures using a Monte Carlo method (Pozdnyakov, Sobol & Sunyaev 1977; 1979; 1983; Górecki & Wilczewski 1984). Analytic corrections to the ST80 calculations at mildly relativistic temperatures and low optical depths have recently been given by Titarchuk (1994) and verified using the Monte Carlo method of Hua and Titarchuk (1994, hereafter HT94). The Comptonization of bremsstrahlung in mildly relativistic plasmas including pairs was treated extensively by Svensson (1982b; 1984) where the conditions for pair equilibrium (i.e., plasmas where the pair production rate equals the annihilation rate) were analytically obtained. In a subsequent series of papers, Zdziarski (1984; 1985, hereafter Z85; 1986) calculated the emergent spectra from pair balanced plasmas including both bremsstrahlung and synchrotron soft photons using a Monte Carlo simulation. Analytic representations of the photon spectra were presented.

In the present analysis we calculate a grid of solutions for mildly relativistic ($\Theta = kT/m_ec^2 \lesssim 2$) plasmas in pair equilibrium using an iterative Monte Carlo procedure. We present new formulas for the thermal bremsstrahlung Gaunt factors at these temperatures and use the exact integral expression for thermal annihilation. Moreover, we calculate the pair opacity due to photon-photon interactions using the the angle-dependent photon distribution rather than through an angle-averaged approximation. We allow for the presence of a soft photon source in addition to the intrinsic bremsstrahlung and annihilation radiation of the plasma. This analysis differs from past numerical simulations of plasmas in pair balance which considered only bremsstrahlung (Zdziarski 1984; McKinley 1986; Liang & Dermer 1987) or thermal cyclo-synchrotron (Zdziarski 1986; Kusunose & Takahara 1985) soft photon sources. We consider, however, only steady-state pair plasmas (see, e.g., Kusunose 1987 for a treatment of the dynamics of thermal pair plasmas). In the pair-free case we compare our results with those of ST80, Z85 and HT94 and specify the limits of applicability for the use of these formulae. The pair balanced solutions that we obtain are presented in the spectral index/compactness plane where direct comparisons with data from high energy observations of Seyferts are made.

In §2 we discuss the bremsstrahlung and pair annihilation radiation processes intrinsic to mildly relativistic thermal plasmas, and in Appendix A we present formulae for the thermal bremsstrahlung Gaunt factors valid at mildly relativistic temperatures. In §3 we describe our Monte Carlo simulations, and in §4 we give the results for pair-free plasmas and compare them with analytic calculations found in the literature. Our results for pair balanced plasmas are given in §5 and compared with data from Seyfert galaxies in §6. We summarize our results in §7.

## 2. Radiation and Pair Balance in Mildly Relativistic Plasmas

In a thermal plasma consisting of electrons, positrons and protons at mildly relativistic temperatures ($\Theta \lesssim 2$), the dominant processes are Compton scattering, bremsstrahlung, photon-photon pair production, and electron-positron annihilation. Higher order processes such as photon-particle and particle-particle pair production, double Compton scattering, and three photon annihilation are negligible at these temperatures (Svensson 1984) and will not be included in this analysis. In addition to the intrinsic radiative processes of the plasma we include, for generality, a soft photon source. Such a soft photon source might arise, for example, if the plasma is threaded with a magnetic field giving rise to thermal self-absorbed synchrotron emission (e.g., Zdziarski 1986). Other sources of soft photons could involve external radiation fields from accretion disks. For optically-thick plasmas the spectrum of the emergent radiation will be modified by transport effects. Absorption due to the inverse bremsstrahlung process can be important for sufficiently low photon energies. In addition, if the plasma has a finite optical depth to Compton scattering, then the scattering of these photons with thermal electrons and positrons will play an important role in determining the emergent spectrum.

The calculation of the photon production due to binary collisions involves integrations of the relavent cross sections over the thermal particle momentum distributions. General integral expressions for photon reaction rates in relativistic plasmas have been derived by Weaver (1976). Formulas for the spectrum of thermal $e^+e^-$ annihilation radiation have been derived by Svensson (1983) and Dermer (1984; for a numerical treatment, see Ramaty & Meszaros 1981). The photon production rate per unit volume per unit energy due to thermal pair annihilation is given by

$$S_a(x;\Theta) = \frac{n_- n_+}{\Theta K_2^2(\frac{1}{\Theta}) m_e c} \exp\left\{-\frac{1}{\Theta}\left[x + \frac{1}{2x}\right]\right\} \int_1^\infty d\gamma_r (\gamma_r - 1) \exp\left(-\frac{\gamma_r}{2x\Theta}\right) \sigma_a(\gamma_r), \tag{2-1}$$

where $x = \epsilon/m_e c^2$ is the dimensionless photon energy in units of the electron rest mass, $K_2(\frac{1}{\Theta})$ is the modified Bessel function of the second kind, and $\gamma_r = \gamma_+\gamma_-(1 - \vec{\beta}_+ \cdot \vec{\beta}_-)$ is the invariant Lorentz factor associated with the relative velocity of the $e^+e^-$ system. The terms $n_-$ and $n_+$ represent the volume densities of electrons and positrons, respectively, and $\sigma_a(\gamma_r)$ is the annihilation cross section, given by

$$\sigma_a(\gamma_r) = \frac{\pi r_e^2}{\gamma_r + 1}\left\{\left(\frac{\gamma_r^2 + 4\gamma_r + 1}{\gamma_r^2 - 1}\right)\ln[\gamma_r + (\gamma_r^2 - 1)^{1/2}] - \frac{\gamma_r + 3}{(\gamma_r^2 - 1)^{1/2}}\right\} \tag{2-2}$$

(e.g., Jauch & Rohrlich 1976). Here $r_e = e^2/m_e c^2 = 2.82 \times 10^{-13}$ cm is the classical radius of the electron.

Relativistic $e^\pm p$, $e^\pm e^\pm$ and $e^+e^-$ thermal bremsstrahlung was calculated by Dermer (1986) using the lowest order bremsstrahlung cross sections. Here we give a semi-analytic representation of the emissivities in the form

$$S_{12}(x;\Theta) = n_1 n_2 \Theta^{-1/2} x^{-1} \exp(-x/\Theta) g_{12}(x;\Theta), \tag{2-3}$$

where we fit various functional forms for the Gaunt factor $g_{12}(x;\Theta)$ to the curves obtained by numerically integrating the exact expressions given by Dermer (1986). Fits to the Gaunt factors in the transrelativistic temperature regime accurate to better than 8% are given in Appendix A, along with asymptotic forms for bremsstrahlung soft photon production at nonrelativistic and ultrarelativistic temperatures (see references in Svensson 1984).

For simplicity we consider a pure hydrogen plasma with spherical geometry and uniform matter density. We consider only plasmas in pair balance and assume that pair escape is negligible. Note that the pair balance condition is nonlinear, insofar as the photon opacity for pair creation is supplied by the photons themselves, which are generated in turn by Compton scattering and annihilation of the electrons and pairs. The fundamental input parameters of the problem are the temperature $\Theta$, the radius $R$, the soft photon luminosity $L_s$, the proton density $n_p$, and the positron density $n_+$. The condition of pair equilibrium is used to iteratively determine $n_+$, and the electron density is fixed by the requirement of charge neutrality, $n_- = n_p + n_+$. Since the reaction rates and luminosities of binary collision processes between any two species with densities $n_1$ and $n_2$ scale as $n_1 n_2 R^3$, it is more convenient to work with the reduced set of parameters $\Theta$, $z = n_+/n_p$, the Thomson depth to ionization electrons $\tau_p = n_p \sigma_T R$, and the soft photon compactness $\ell_s = L_s \sigma_T / R m_e c^3$, where $\sigma_T = 6.652 \times 10^{-25}$ cm$^2$ is the Thomson cross section. The total Thomson optical depth $\tau = (1 + 2z)\tau_p$. In terms of these variables, $R$ and $n_p$ enter only through the product $\tau_p = n_p \sigma_T R$. The luminosity of the emergent radiation scales as $R$ for a given value of $\tau_p$, whereas the shape of its spectrum is unchanged. This scaling holds whenever most of the soft photon energy density is injected at frequencies well above the free-free self absorption frequency. This is the case for dimensionless soft photon energies $x \gtrsim 5 \times 10^{-8} \tau \Theta^{-3/4}$, corresponding to the free-free self absorption frequency for an emission region of size $R = 3 \times 10^7$ cm, which corresponds to 10 Schwarzschild radii for a 10 $M_\odot$ black hole.

## 3. Description of the Monte Carlo Code

We calculate the cumulative effects of Compton scattering and $\gamma\gamma$ pair attenuation self-consistently using an iterative Monte Carlo approach. When a photon is Compton scattered, the final energy and angle through which it scatters are randomly drawn from the distributions determined from the Klein-Nishina cross section and the relativistic Maxwellian distribution of the electrons. This is done by first performing a Monte Carlo simulation of single Compton scattering of a photon with an electron in the thermal plasma, and constructing scattering energy and angle arrays. These arrays are used in a second program, described below, which follows the photon propagation in the hot plasma.

In the program that generates the scattering arrays, a given photon with specified initial energy $x$ is scattered by an electron chosen with a random direction and energy $\gamma m_e c^2$ drawn from a relativistic Maxwellian of temperature $\Theta$. If the photon scatters through an angle cosine of $\mu_s$ then the scattered photon will have the energy given by (e.g., Pozdnyakov et al. 1983)

$$x_s = \frac{x(1 - \beta\mu_i)}{1 - \beta\mu_f + (x/\gamma)(1 - \mu_s)}, \tag{3-1}$$

where $\mu_i$ and $\mu_f$ are the cosines of the angles between the initial electron velocity vector $\vec{\beta}$ and the initial and final photon propagation directions, respectively. The angle through which the photon scatters, $\mu_s$, is given by

$$\mu_s = \mu_i \mu_f + [(1 - \mu_i^2)(1 - \mu_f^2)]^{1/2} \cos(\phi_f - \phi_i), \tag{3-2}$$

where $\phi_i$ and $\phi_f$ are the azimuthal photon directions. This scattering event is assigned a weight given by

$$w(x, x_s, \mu_s) = (1 - \beta\mu_i) \frac{d\sigma}{d\Omega_s}. \tag{3-3}$$

where the polarization-averaged Klein-Nishina cross section for the scattering of a photon with a moving electron is (e.g. Pozdnyakov et al. 1983)

$$\frac{d\sigma}{d\Omega_s} = \frac{r_e^2}{2}\frac{1}{\gamma^2(1-\beta\mu_i)^2}\left(\frac{x_s}{x}\right)^2\left[\frac{x^*}{x_s^*}+\frac{x_s^*}{x^*}+2\left(\frac{1}{x^*}-\frac{1}{x_s^*}\right)+\left(\frac{1}{x^*}-\frac{1}{x_s^*}\right)^2\right], \tag{3-4}$$

with

$$x^* = \gamma(1-\beta\mu_i)x, \tag{3-5}$$

and

$$x_s^* = \gamma(1-\beta\mu_f)x_s. \tag{3-6}$$

These weights are summed in bins of scattered photon energy, $x_s$, and scattering angle cosine, $\mu_s$, for a grid of incident photon energies. In this way we obtain the probability $P(x,\mu_s,x_s)dx_s d\mu_s$ for the scattering of a photon with energy $x$ into the angle cosine range $d\mu_s$ about the angle $\cos^{-1}\mu_s$ and into the energy range $dx_s$ around $x_s$ by an electron in a plasma of temperature $\Theta$. We then tabulate the cumulative distributions given by

$$F_x(x_s) = \int_0^{x_s} dx_s' \int_{-1}^{1} d\mu_s'\, P(x,\mu_s',x_s'), \tag{3-7}$$

and

$$G_{x,x_s}(\mu_s) = \int_{-1}^{\mu_s} d\mu_s'\, P(x,\mu_s',x_s). \tag{3-8}$$

The energy of the scattered photon, $x_s$, and the scattering angle, $\mu_s$, can be obtained from two uniform random numbers, $\xi,\eta \in [0,1]$ by inverting the cumulative probability distribution functions:

$$x_s = F_x^{-1}(\xi), \tag{3-9}$$

$$\mu_s = G_{x,x_s}^{-1}(\eta). \tag{3-10}$$

For a given simulation we first specify the input parameter set

$$P_I = \{\Theta,\ \tau_p,\ z,\ l_s\}. \tag{3-11}$$

Photons are then injected into the plasma sphere with random direction and specified spatial distribution. We assume a uniform distribution except when we make comparisons with the results of Sunyaev and Titarchuck (1980), in which case we use the distribution they employed to obtain analytic solutions in the nonrelativistic diffusion regime. This distribution is given by

$$f(r) = \frac{R}{\pi r}\sin\left(\frac{\pi r}{R}\right), \tag{3-12}$$

where $r$ is the distance from the center of the sphere. The energy of the injected photons is drawn from the bremsstrahlung, annihilation and soft photon emissivities. For the soft photons we assume either a black body spectrum with temperature $kT = 0.1$ keV or a mono-energetic distribution at energy 0.33 keV. The relative abundances of bremsstrahlung, annihilation and soft photons are governed by the input parameter set $P_I$.

Although bremsstrahlung absorption is generally unimportant, we determine the relative probability that a photon is scattered, absorbed by the inverse bremsstrahlung process, or escapes from the system. If the photon is absorbed we exit the loop and choose a new photon. If the photon is scattered, a new photon energy and direction are obtained and the process is repeated until either the photon escapes or is lost by free-free absorption. Those that escape are binned in energy, building up the emergent spectrum. At each step the energy and direction of the photon and the radial location of the scattering event are recorded. From this data we obtain the photon density $n^{(0)}(r, \mathbf{\Omega}, x)$.

In the first iteration $\gamma\gamma$ pair production is neglected. The simulation is then repeated with the inclusion of an approximation to the energy-dependent radially-averaged $\gamma\gamma$ opacity, $\alpha^{(1)}_{\gamma\gamma}(x)$, obtained from the photon density distribution $n^{(0)}(r, \mathbf{\Omega}, x)$ established in the first iteration. First, the optical depth $\tau^{(0)}_{\gamma\gamma}(x)$ along the radius of the sphere for a photon of energy $x$ is obtained by numerically evaluating the integral

$$\tau^{(1)}_{\gamma\gamma}(x) = \int_0^R dr \oint d\mathbf{\Omega} \int_{\frac{2}{x(1-\mu)}}^{\infty} dx'\, n^{(0)}(r, \mathbf{\Omega}, x')(1-\mu)\sigma_{\gamma\gamma}(x, x', \mu), \tag{3-13}$$

where $\sigma_{\gamma\gamma}(x, x', \mu)$ is the total cross section for the process $\gamma\gamma \to e^+e^-$ and is given by (e.g. Jauch & Rohrlich 1976)

$$\sigma_{\gamma\gamma}(x, x', \mu) = \frac{1}{2}\pi r_e^2(1-\beta^2)\Big[(3-\beta^4)\ln\Big(\frac{1+\beta}{1-\beta}\Big) - 2\beta(2-\beta^2)\Big]. \tag{3-14}$$

Here $\beta$ is the velocity of either the electron or positron in the center of mass and is related to $x$, $x'$ and $\mu$ by the relation

$$\beta = \left[1 - \frac{2}{xx'(1-\mu)}\right]^{1/2}. \tag{3-15}$$

We evaluate the integral (3-13) by the Monte Carlo technique and then set

$$\alpha^{(1)}_{\gamma\gamma}(x) = \tau^{(1)}_{\gamma\gamma}(x)/R. \tag{3-16}$$

We then repeat the entire process, this time including the effects of $\gamma\gamma$ attenuation through $\alpha^{(1)}_{\gamma\gamma}(x)$. This results in a new spectrum and photon density distribution $n^{(1)}(r, \mathbf{\Omega}, x')$ from which we obtain $\alpha^{(2)}_{\gamma\gamma}(x)$ and so forth. For pair-poor plasmas ($z \ll 1$) the first iteration is always sufficient for establishing equilibrium. For pair-dominated plasmas ($z \gtrsim 1$) we find that pair balance is usually obtained at the second or third iteration.

## 4. Pair-Free Solutions: Comparisons With Analytic Results

In this section we give results for the case of a pair-free plasma ($z = 0$) where the soft photon luminosity greatly exceeds that of bremsstrahlung. Thus $\tau = \tau_p$. We inject soft photons with energy $xm_ec^2 = 0.33$ keV, a random propagation direction, and spatial distribution given by equation (3-12) into a sphere of radius $R = 3 \times 10^7$ cm. The choice of this spatial distribution is made in order to compare with the analytic results of ST80. In addition, we compare our results with the analytic bridging formulae of Z85 and the relativistic corrections to the ST80 calculations given by HT94. We set $z = 0$, $R = 3 \times 10^7$ cm and $\ell_s \gg \ell_{ff}$, where the bremsstrahlung soft photon compactness is given by

$$\ell_{ff} \approx 4.3\alpha_f \tau^2 \Theta^{1/2}(1 + 1.2\Theta + 1.8\Theta^{1.34} + 1.3\Theta^2 + 1.2\Theta^3 - 1.5\Theta^{3.5}) \tag{4-1}$$

for $\Theta \leq 1$ (Svensson 1982) and $\alpha_f$ is the fine-structure constant. This reduces the parameter set $P_I$ to $\{\Theta,\tau\}$.

In Figures 1a-1d we compare the results of our Monte Carlo simulations for $\delta$-function soft photon injection with the analytic results of ST80, Z85 and HT94 for the teperatures $\Theta = \{0.1, 0.2, 0.5, 1.0\}$ and Thomson depths $\tau = \{0.1, 0.5, 1.0, 5.0\}$. All curves are normalized such that the integral over energy is unity. As can be seen, in the high optical depth, low temperature regime of thermal Comptonization, all analytic fits are in good agreement with the Monte Carlo simulation. This is expected because the analytic fits are designed to approach the Sunyaev-Titarchuk result derived in the large optical depth diffusion limit. As $\tau$ becomes much less than unity, the ST80 results break down whereas the HT94 analytic fit still provides a good representation, because they modify the escape probability used in the Sunyaev-Titarchuk limit to apply to the low optical depth case. The fit of Z85 is poor in the low temperature, low $\tau$ case because he is only considering parameters leading to photon spectral indices softer than $\Gamma = 1.4$, which requires large optical depths in the low temperature limit. At temperature $\Theta = 0.2$ (Fig. 1b), both Z85 and HT94 are in good agreement with the Monte Carlo results at large optical depths, although Z85 is again discrepant at low optical depths. At still higher temperatures shown in Fig. 1c and 1d, serious discrepancies appear between the analytic fits and the simulation results.

We, therefore, recommend the use of the model of Sunyaev and Titarchuk (1980) for $\tau \gtrsim 2$ and $\Theta \lesssim 0.2$. The corrections of Titarchuk (1994) (see also Hua & Titarchuk 1994) extend the validity of this model to $\tau \gtrsim 0.2$ and $\Theta \lesssim 0.3$. If the photon spectral index is in the range $1.4 \lesssim \Gamma \lesssim 2.5$, which occurs for small $\tau$ and large $\Theta$ or, alternatively, small $\Theta$ and large $\tau$, then the representation of Zdziarski (1985) is in agreement with our simulations. We advise caution in the use of these formulas when $\tau \gtrsim 1$ and $\Theta \gtrsim 1$ because at these temperatures and optical depths, the solutions may not be in pair balance and bremsstrahlung contributes significantly, a situation we consider in more detail in §5.

In Figure 2 we show the variation of the 2-10 keV photon number spectral index, $\Gamma$(2-10 keV), as a function of $\tau$ for the temperatures $\Theta = \{0.1, 0.2, 0.5, 1.0\}$. We again see good agreement in the large optical depth limit, but large discrepancies are found in the low optical depths for the analytic fits that are restricted to the diffusion limit (ST80) or to spectra that are relatively hard (Z85). The results of HT94 are in good agreement except at high temperatures. In the low optical depth limit of the $\Theta = 1$ curve, however, the Monte Carlo simulations are very sensitive to the injection energy of the photons because the amplification factor is large and the Compton-scattered spectrum mirrors the effects of the individual scattering events (see Pozdnyakov et al. 1983). This effect is shown clearly in Figure 3, where we plot values of $\Gamma$(2-10 keV) as a function of the temperature for the blackbody soft photon distribution scattered by thermal electrons with $\Theta = 0.5$. When $\tau = 0.01$, we see variations of $\Gamma$ by nearly 2 units.

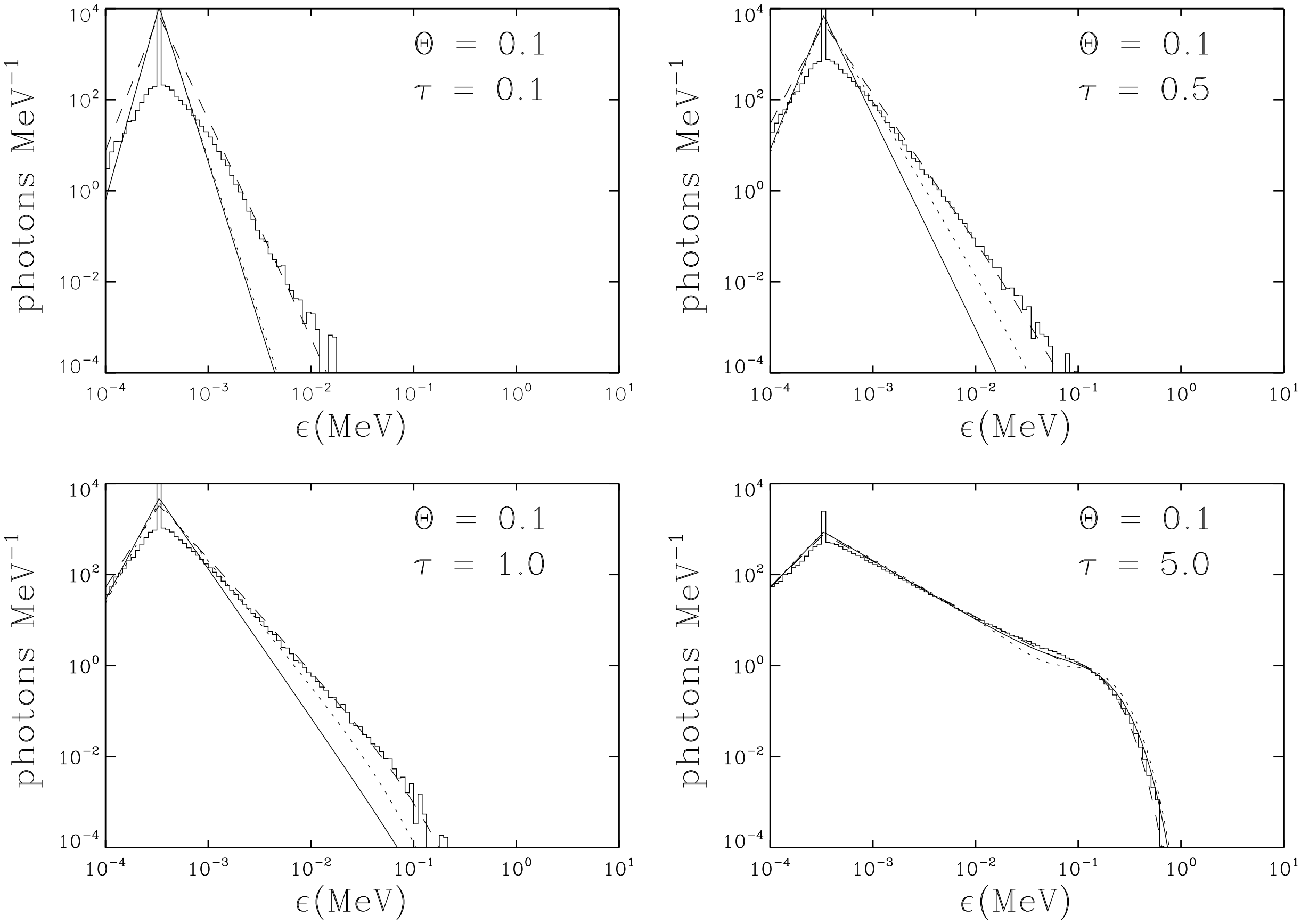

Fig. 1a.— The emergent photon spectra resulting from Monte Carlo simulations of the Comptonization of monochromatic soft photons with energy 0.33 keV (histograms) compared with the analytic approximations of Sunyaev & Titarchuk (1980; solid curves); Zdziarski (1985; dotted curves), and Hua & Titarchuk (1994; dashed curves). The dimensionless electron temperature $\Theta = 0.1$, and spectra for different optical depths of electrons are shown by the labels in the separate panels.

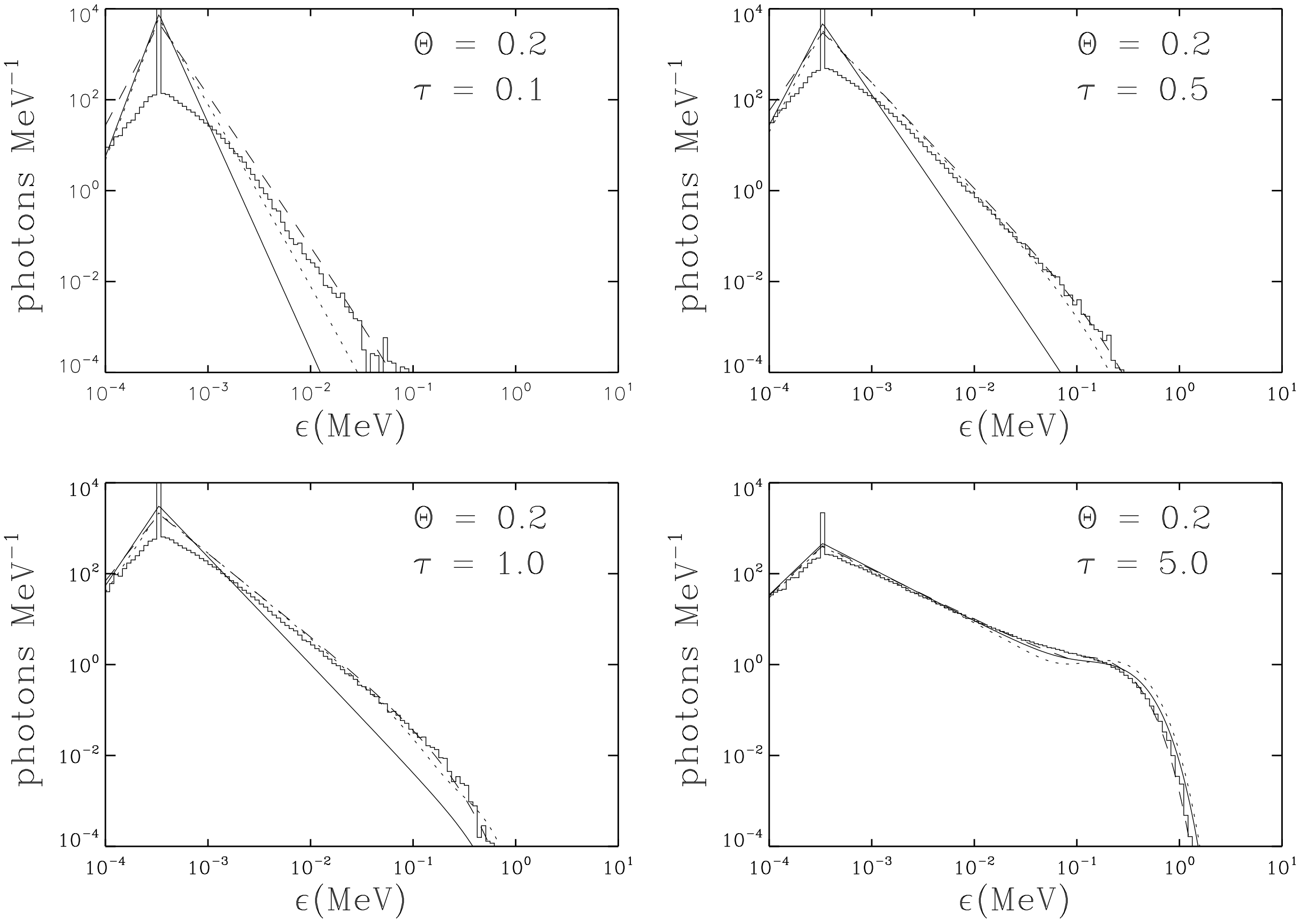

**Fig. 1b.**— Same as for Fig. 1a, but with $\Theta = 0.2$.

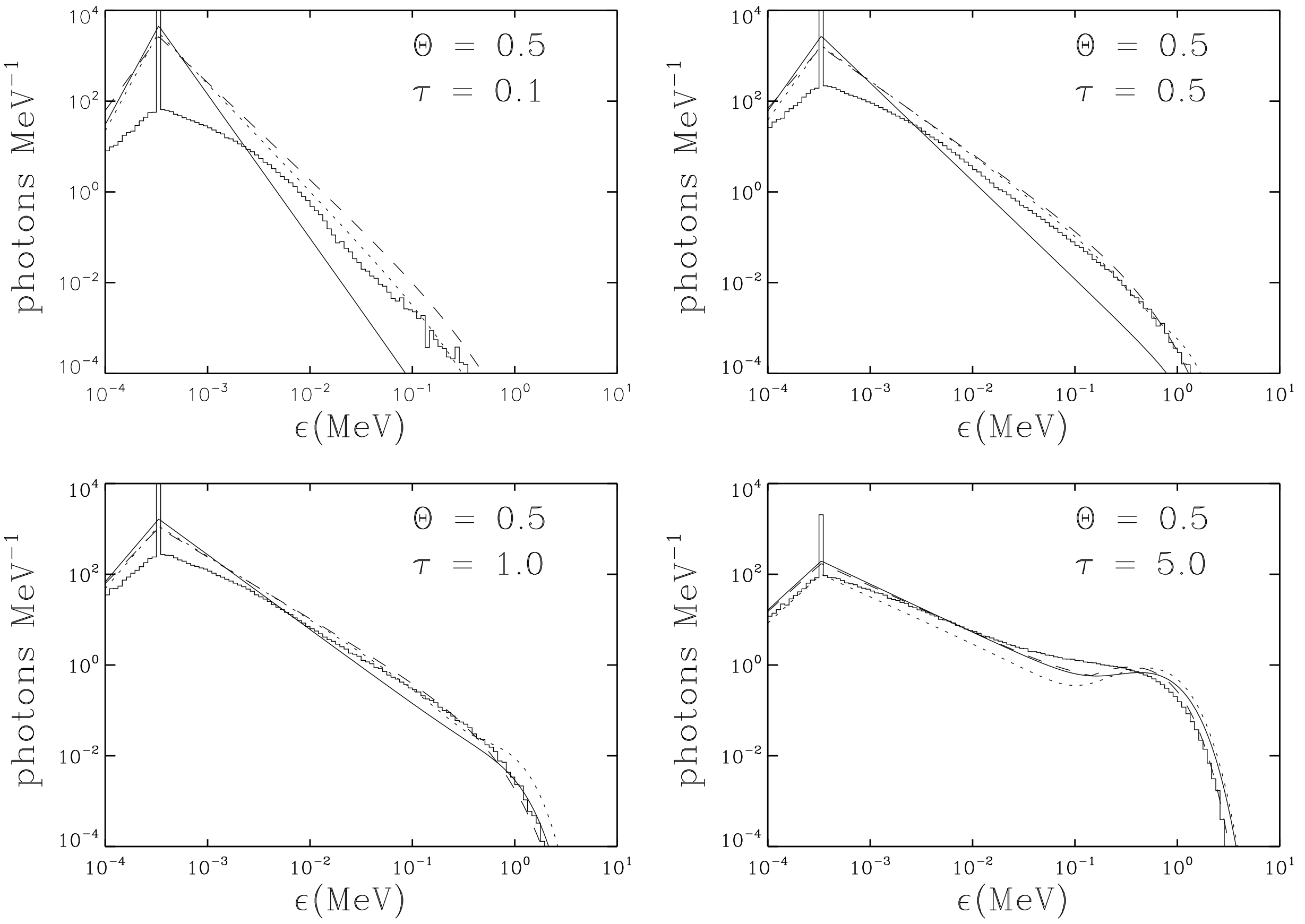

Fig. 1c.— Same as for Fig. 1a, but with $\Theta = 0.5$.

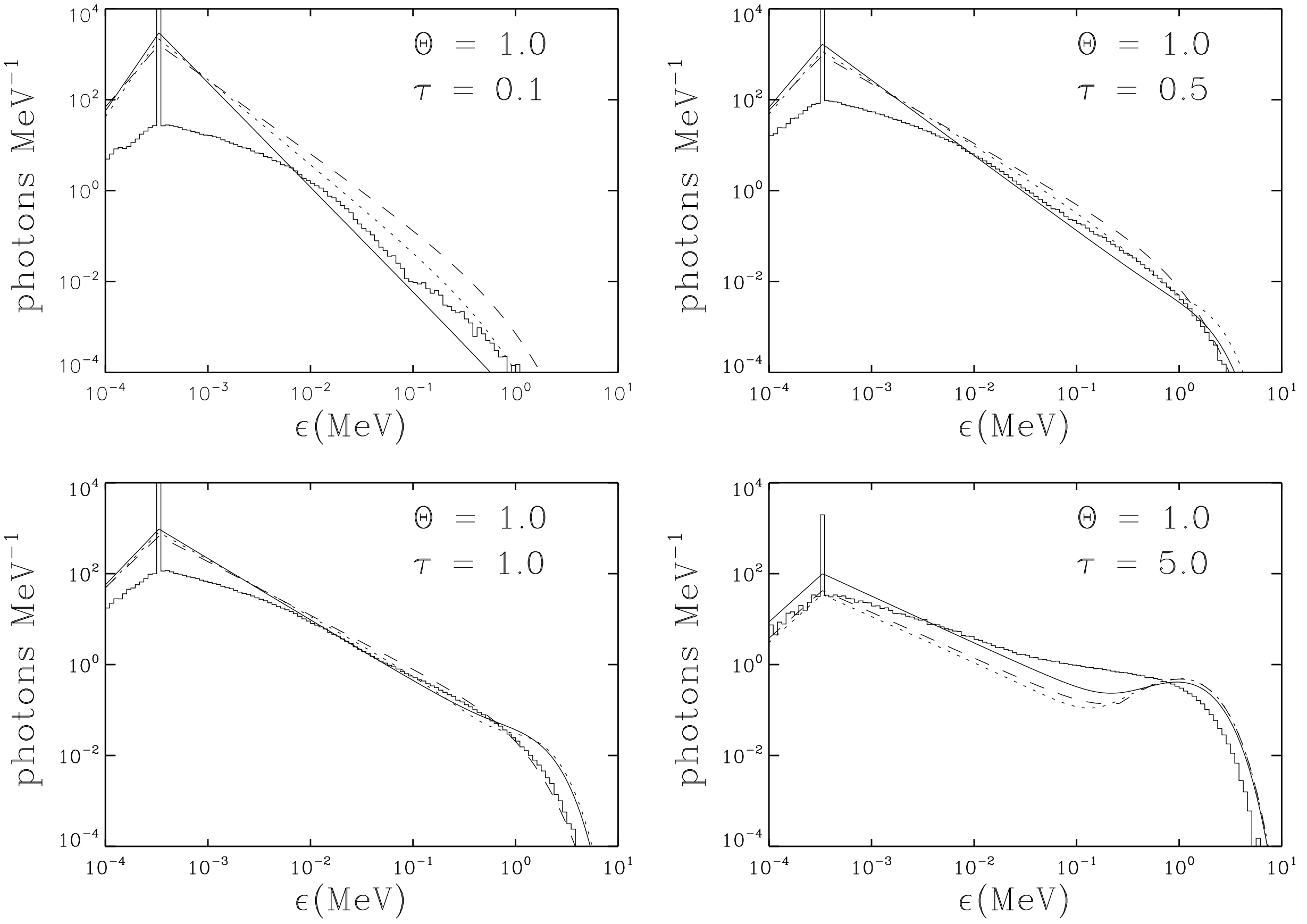

Fig. 1d.— Same as for Fig. 1a, but with $\Theta = 1.0$.

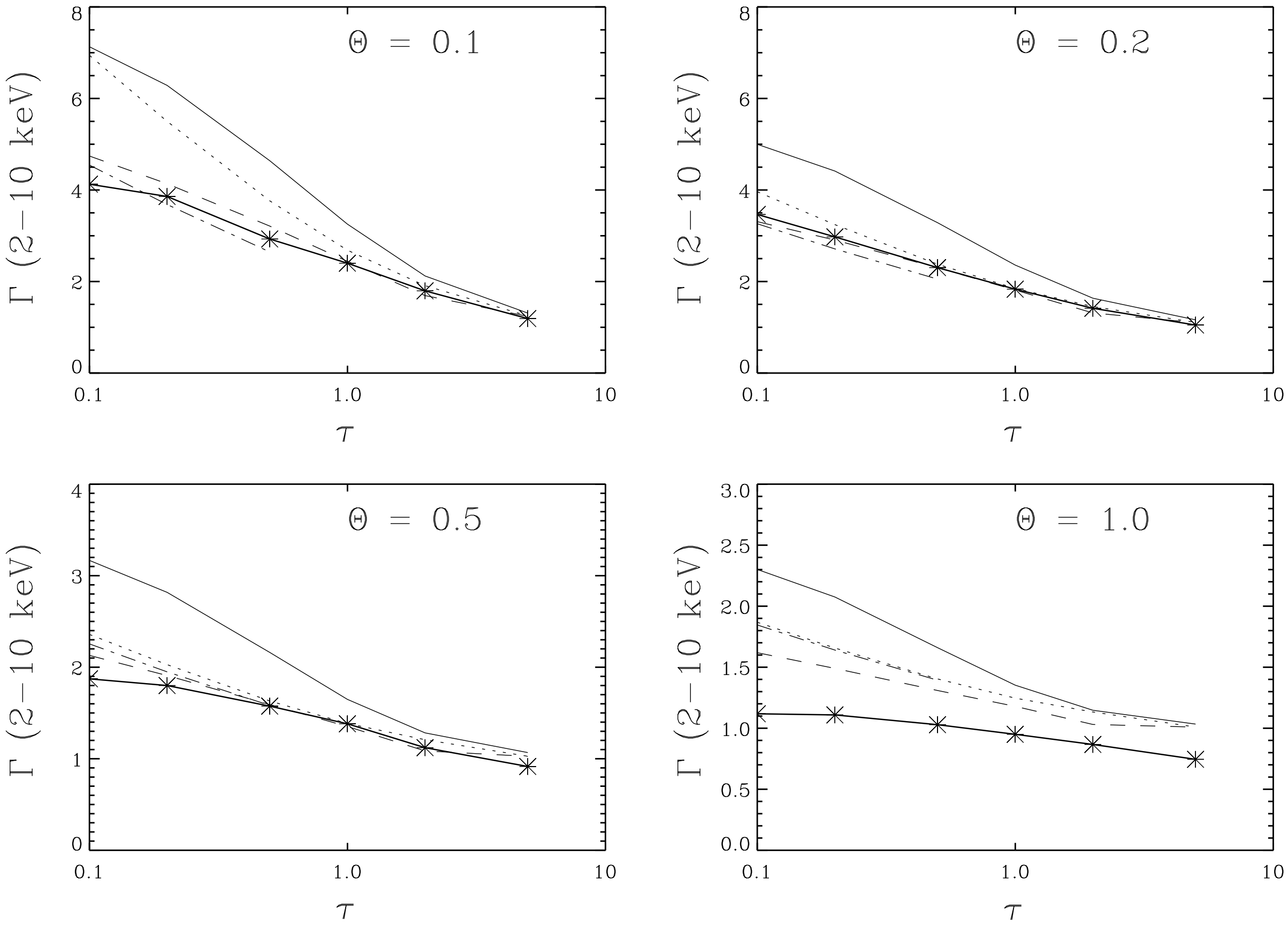

Fig. 2.— Monte Carlo calculations of the 2-10 keV photon spectral index resulting from thermal Comptonization of monochromatic photons with energy 0.33 keV (heavy solid curves) are compared with the analytic approximations of Sunyaev & Titarchuk (1980; solid curves); Zdziarski (1985; dotted curves), and Hua & Titarchuk (1994; dashed curves). Also shown by the dot-dashed curves are the low temperature, low optical depth approximations to the spectral index given by equations (4-2) with equation (4-5) replacing equation (4.3).

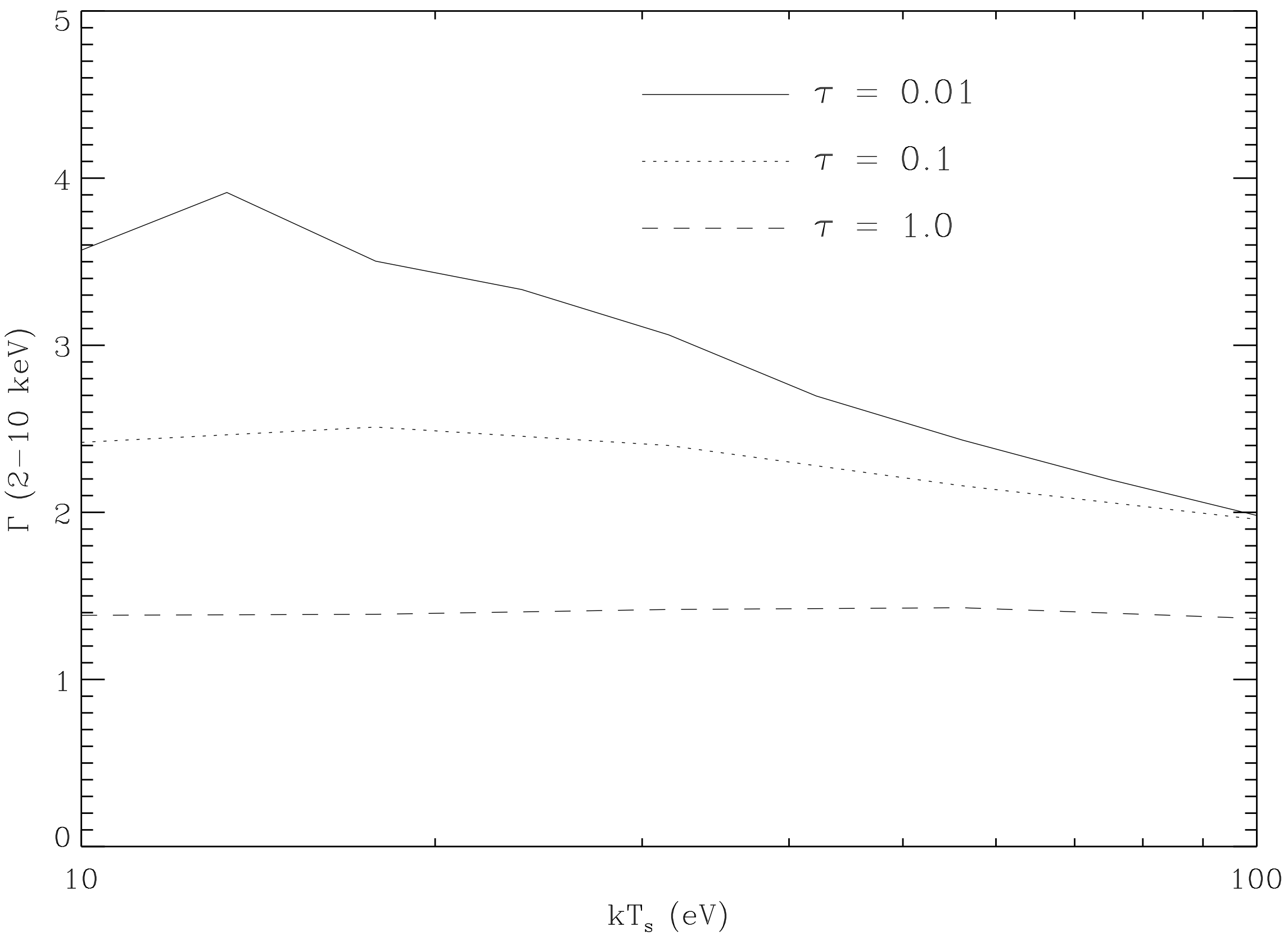

Fig. 3.— The variation of the 2-10 keV photon spectral index with the temperature of the soft photon blackbody for $\tau$=0.01, 0.1 and 1.0.

In the low temperature, low optical depth regime, the spectral index is harder than expected using arguments by Zel'dovich (see Pozdnyakov, Sobol', & Sunyaev 1977; Zdziarski 1986; Dermer, Liang, & Canfield 1991), which imply that the photon index

$$\Gamma = 1 - \ln P(\tau)/\ln A. \tag{4-2}$$

Here the energy amplification factor

$$A(\Theta) = 1 + \frac{4\Theta K_3(1/\Theta)}{K_2(1/\Theta)} \cong 1 + 4\Theta + 16\Theta^2 \tag{4-3}$$

is the ratio of the mean photon energy after scattering to its initial energy in the limit $x \to 0$ and

$$P(\tau) = 1 - \frac{3}{8\tau^3}[(2\tau^2 - 1) + e^{-2\tau}(2\tau + 1)] \tag{4-4}$$

is the mean scattering probability in a uniform sphere (see Osterbrock 1974).

We find that the formation of the spectrum in the low temperature, low optical depth regime is controlled by the spread of the photon number distribution rather than through the average increase in energy of a photon after scattering. Approximating a nonrelativistic thermal electron distribution by a monoenergetic distribution with speed $\beta = (3\Theta)^{1/2}$, one finds that with each scattering, the photon distribution spreads by a factor $\approx 1 + (3\Theta)^{1/2}$. Using analogous arguments to those given by Zel'dovich, this implies a spectral index in this regime given by equation (4-2), but with the amplification factor $A$ replaced by a new factor $A' \approx 1 + (3\Theta)^{1/2}$, representing the spread of the photon distribution function. We can generalize the spectral index approximation (4-2) to arbitrary temperatures by replacing $A(\Theta)$ with the new factor

$$A^*(\Theta) \cong 1 + (3\Theta)^{1/2} + 4\Theta + 16\Theta^2. \tag{4-5}$$

Good agreement with the numerical results is obtained in the low temperature, low optical depth regime with this replacement. Figure 2 shows a comparison of the $\tau \to 0$ approximation using equations (4-2) with equation (4-5) replacing equation (4-3).

## 5. Pair Balanced Solutions

In this section we give results for the pair-balanced solutions obtained with our Monte Carlo simulations. In addition to the intrinsic bremsstrahlung and annihilation radiation of the plasma, we inject soft photons with a blackbody energy distribution at temperature $kT = 0.1$ keV and soft-photon compactness $\ell_s$. We vary the parameter set $P_I = \{\Theta, \tau_p, z, \ell_s\}$ by first fixing $\Theta$, $\tau_p$ and $\ell_s$ and then adjusting $z$ until pair balance is achieved. There are either zero, one or two pair-balanced states for a given $\Theta$, $\tau_p$ and $\ell_s$ (Svensson 1984; Zdziarski 1984, 1985). A low $z$ solution occurs when the photon density at energies above threshold for pair production is low enough such that the small amount of pair production that ensues can be balanced by the steady state annihilation of a relatively small amount of pairs. As the pair fraction $z$ is increased, the annihilation rate will at first exceed the production rate until the scattering opacity due to the pairs becomes comparable to that of $\tau_p$. At this point the spectrum will harden as more low energy photons get Comptonized to higher energies. The pair production increases in a nonlinear fashion until balance is again obtained. This is the high $z$, pair dominated regime.

In Figure 4 we show the results for $\ell_s = 0$. The compactness of the emergent radiation $\ell_h$ is plotted as a function of $\Theta$, where each curve represents a different value of $\tau_p$. We compare our results with the analytic calculations of Svensson (1984) and the Monte Carlo simulations of Zdziarski (1985) for the case $\tau_p = 1$. Factor-of-two deviations between our numerical results and the analytic treatment can be attributed to the uncertainty in deriving the number of scatterings required for a soft photon to enter the Wien regime and pair produce. We find good agreement between our results and the calculations of Zdziarski in the regime where $z \ll 1$, which occurs on the lower branches of the curves in Figure 4. Small discrepancies between the numerical simulations in the upper pair-dominated portion of the curves are due to an improved expression for the $e^+e^-$ bremsstrahlung emissivity and a different method for calculating $\tau_{\gamma\gamma}$. In this work we used the full angular distribution of the photons to calculate the pair opacity, whereas the photon distribution was assumed to be isotropic in the analysis of Zdziarski (1984).

Figure 5 shows the effects of increasing the soft photon compactness $\ell_s$ on the allowed states of a thermal plasma in pair balance. Here we plot the hard compactness $\ell_h$ as a function of $\Theta$ for different values of $\tau_p$ and $\ell_s$. The outermost curves in the panel correspond to the $\ell_s = 0$ solutions of Figure 4. The effect of increasing the soft photon compactness is to increase the hard compactness and to decrease the maximum temperature $\Theta_{max}(\ell_s, \tau_p)$ permitted for a steady pair-balanced plasma.

In Figure 6 we show examples of spectra resulting from our simulations. These spectra correspond to plasmas in pair-balance with no external soft photon injection ($\ell_s = 0$) and the parameter values $\Theta$=$\{0.2, 1.0\}$ and $\tau_p$=$\{0.1, 1.0\}$. For each set of parameters we show the spectra for the two states of the plasma permitted by the condition of pair-balance. As can be seen, in the low-$z$, low optical depth ($\tau_p = 0.1$) states, the plasma simply emits a thermal bremsstrahlung spectrum. The dotted curves for the $z = 0$ cases correspond to pure thermal bremsstrahlung calculated using the Gaunt factor given in Appendix A. The histogram, which includes the effects of Comptonization, shows little or no departure in the $\tau_p = 0.1$ cases. The situation, however, is different for $\tau_p = 1.0$ where the effects of Comptonization are manifest by the exess at high energies. The effects of Comptonization and pair annihilation are clearly seen in the high-$z$ states. Here the dotted curves represent the sum of annihilation and thermal bremsstrahlung.

The Comptonized spectra appear harder over most of the energy range and saturate at energies near $3\Theta$ to form a Wien bump. In addition, the difference between the hard compactnesses of the high-$z$ and low-$z$ states decreases with increasing $\Theta$ and $\tau_p$ until the two solutions merge as shown in Figure 4. This is apparent in Figure 6 by examining the overall levels of the emission of the two states.

In Figure 7 we show the effect of injecting photons with a uniform spatial distribution and a blackbody energy distribution at $kT = 0.1$ keV on the emergent spectra for the high-$z$ pair-balanced states of Figure 6. The presence of the soft photon graybody is clearly visible and the individual histograms can be identified with values given for $\ell_s$ in each panel according the relative levels of the graybody peak. The histograms displaying no soft graybody ($\ell_s = 0$) correspond to the high-$z$ solutions of Figure 6. In each case the soft photon compactness is increased to roughly the maximum value permitted by the condition of pair-balance. It is evident from this figure that for higher $\Theta$ and $\tau_p$ the plasma becomes increasingly unable to accomodate the injection of soft photons. Hence, for $\Theta \gtrsim 1.0$ and $\tau_p \gtrsim 1.0$, steady solutions obtained neglecting the intrinsic thermal bremsstrahlung of the plasma are not physically realistic.

In Figure 8 we plot the isotherms in $\Gamma$-$\ell_h$ space, where $\Gamma$ represents the 2 and 10 keV spectral index here and in the subsequent discussion. Each panel corresponds to a different value of $\tau_p$. The isotherms are generated by fixing $\Theta$ and $\tau_p$, varying the value of the parameter $\ell_s$, and determining the values of $z$ for which the plasma is in pair balance. The symbols in this figure give the numerically calculated solutions, and the endpoints of the isotherms correspond to the $\ell_s = 0$ solutions shown in Figure 4. The different values of $\ell_s$ at each point can be determined from Figure 5. The pair-dominated solutions generally cluster near the high-compactness end of the isotherms, and pair-dominated solutions at a given temperature coincide, independent of $\tau_p$, provided that $\tau_p$ is not too large. This is because the Comptonization of soft photons and the pair balance condition is determined by the Thomson depth $\tau$ which is $\gg \tau_p$ when $z \gg 1$.

We can divide each panel conceptually into three regimes. In the left-hand portion we find the bremsstrahlung regime, where the soft photon energy density is dominated by bremsstrahlung, the solutions have few pairs ($z \ll 1$), and the spectra are hard, corresponding to nearly pure bremstrahlung solutions. This regime represents the solutions for which $\ell_s \lesssim \ell_{ff}$ (see equation 4-1), which are located near or at the left-most endpoints of each curve. As $\ell_s$ becomes $\gg \ell_{ff}$, the spectra soften and reach a plateau where the formation of the spectrum is dominated by non-bremsstrahlung soft-photon Comptonization and $\ell_h \propto \ell_s$. This is the linear regime, which continues with increasing $\ell_s$ until the Compton opacity due to pairs is comparable to that of the ionization electrons. As discussed previously, a second solution with large Compton opacity from pairs is found through the condition of pair balance. The solutions in this pair-dominated regime are harder than the corresponding pair-dilute solutions for a given value of $\ell_s$, and persist even if $\ell_s = 0$ (see Figure 4).

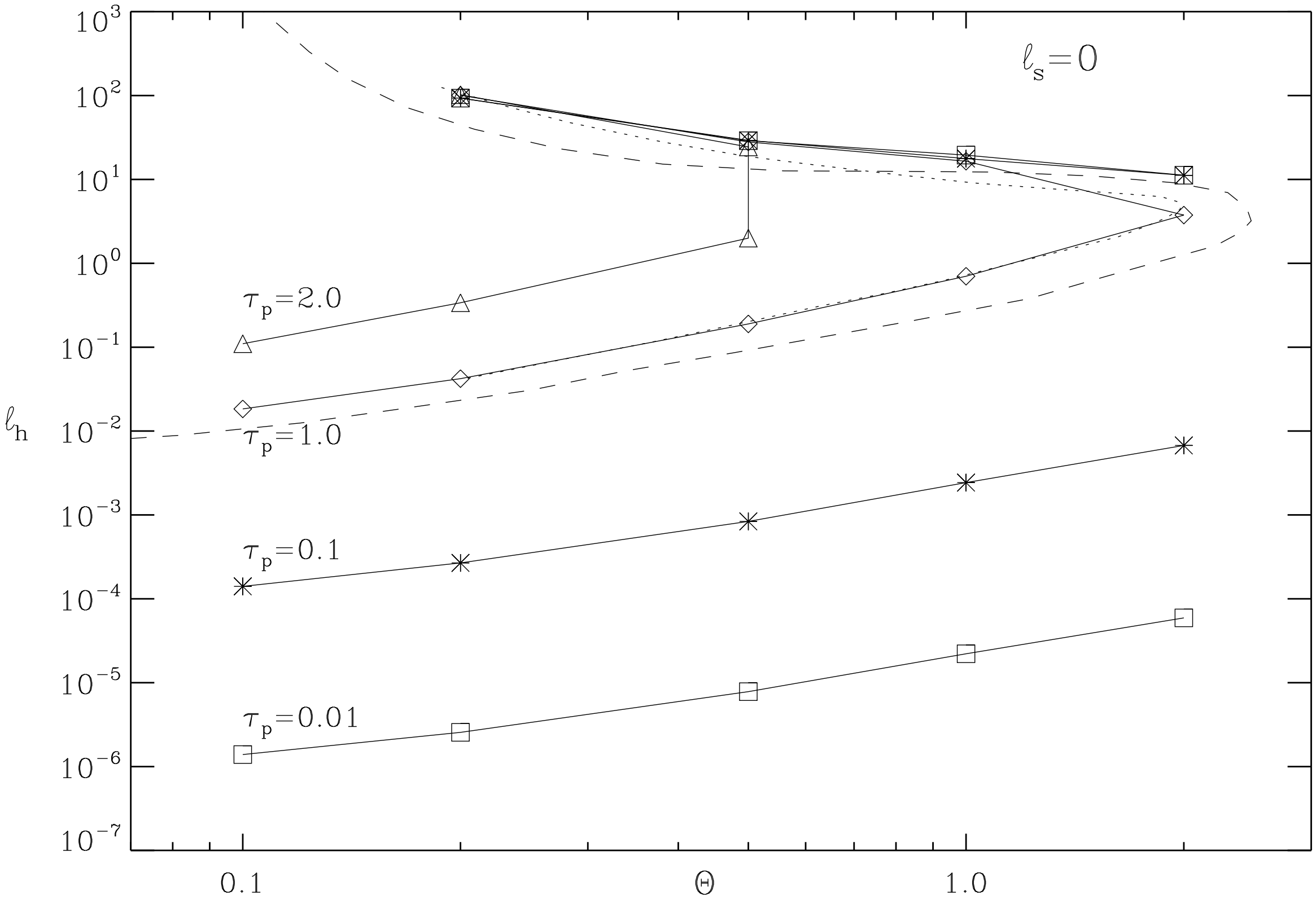

Fig. 4.— Hard compactness $\ell_h$ as a function of $\Theta$ for thermal plasmas in pair balance when all soft photons are supplied by bremsstrahlung ($\ell_s \to 0$). Curves are labeled by different values of the proton optical depth $\tau_p$. The dotted and dashed curves respectively show the numerical result of Zdziarski (1984) and the analytic result of Svensson (1984) for the case $\tau_p = 1$.

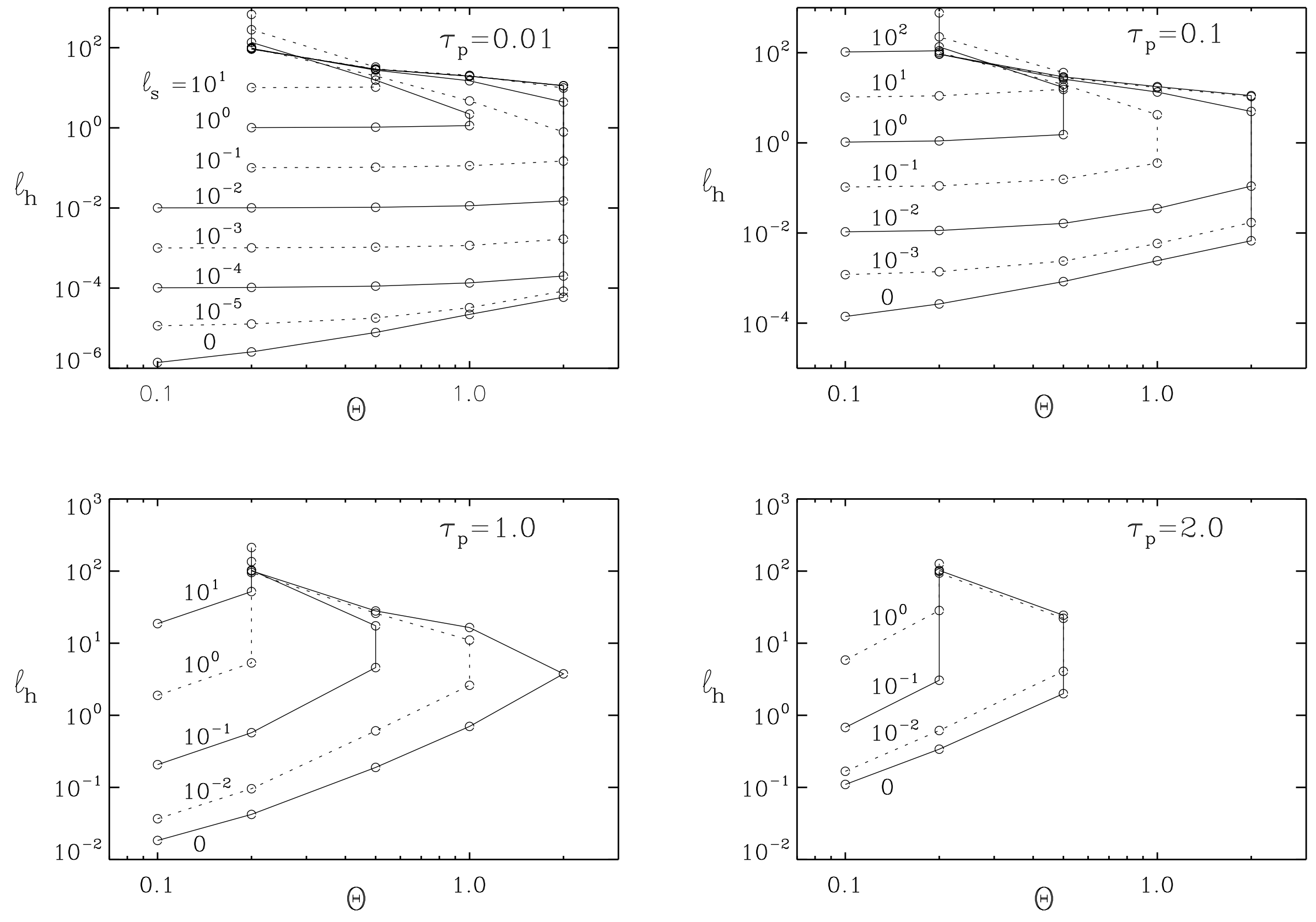

Fig. 5.— Hard compactness $\ell_h$ as a function of $\Theta$ for thermal plasmas in pair balance for soft photons with compactness $\ell_s$ injected uniformly into a spherical cloud with uniformly distributed electrons and pairs. The solutions in the separate panels correspond to different values of the proton optical depth $\tau_p$.

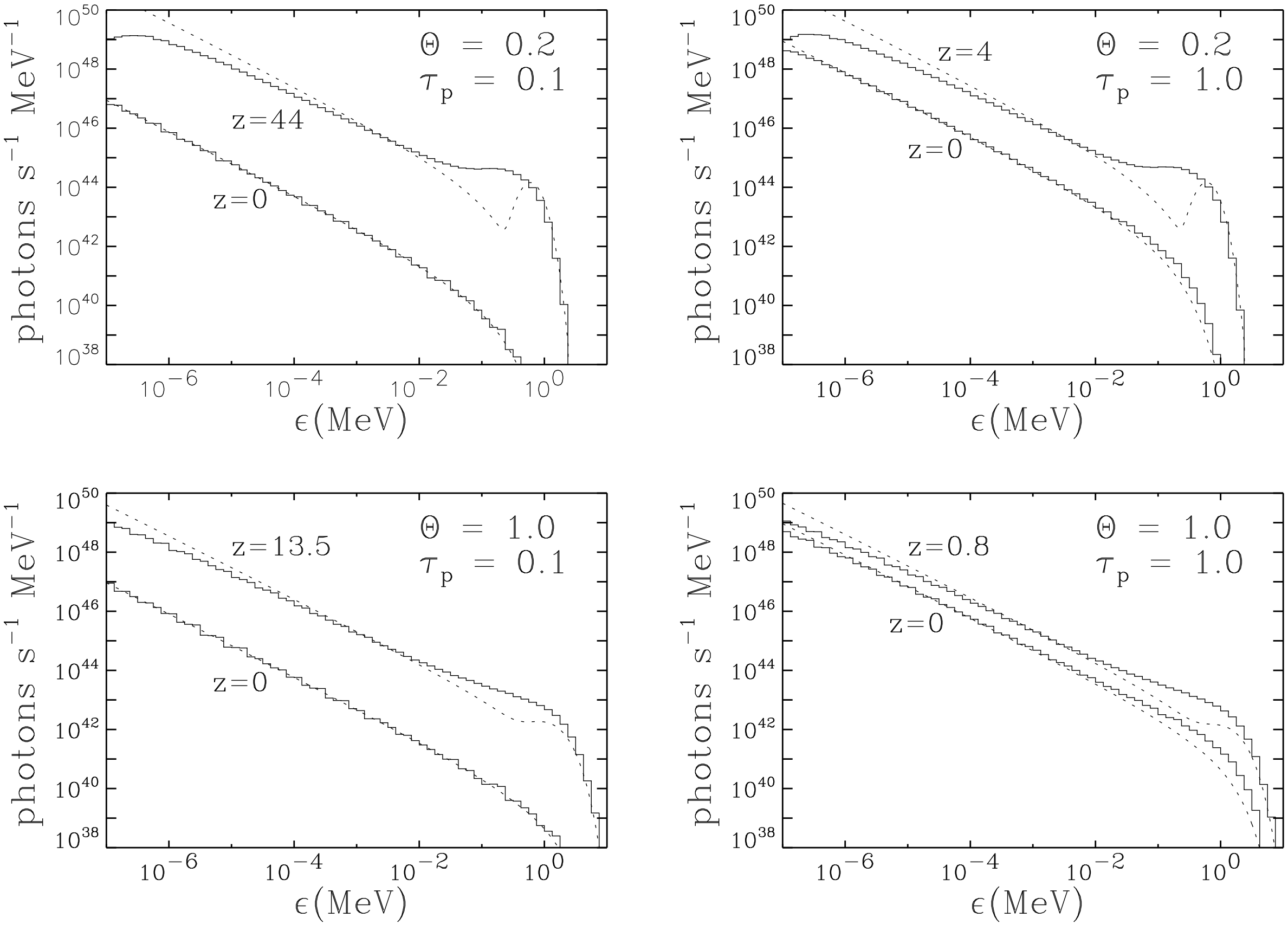

Fig. 6.— Spectra corresponding to the low-$z$ and high-$z$ pair-balanced states of a plasma with the parameters $\Theta = \{0.2, 1.0\}$, $\tau_p = \{0.1, 1.0\}$ and $\ell_s$=0. The dotted curves show the bremsstrahlung and pair annihilation spectra of the plasma without Comptonization. The solid histogram represents the fully Comptonized spectra.

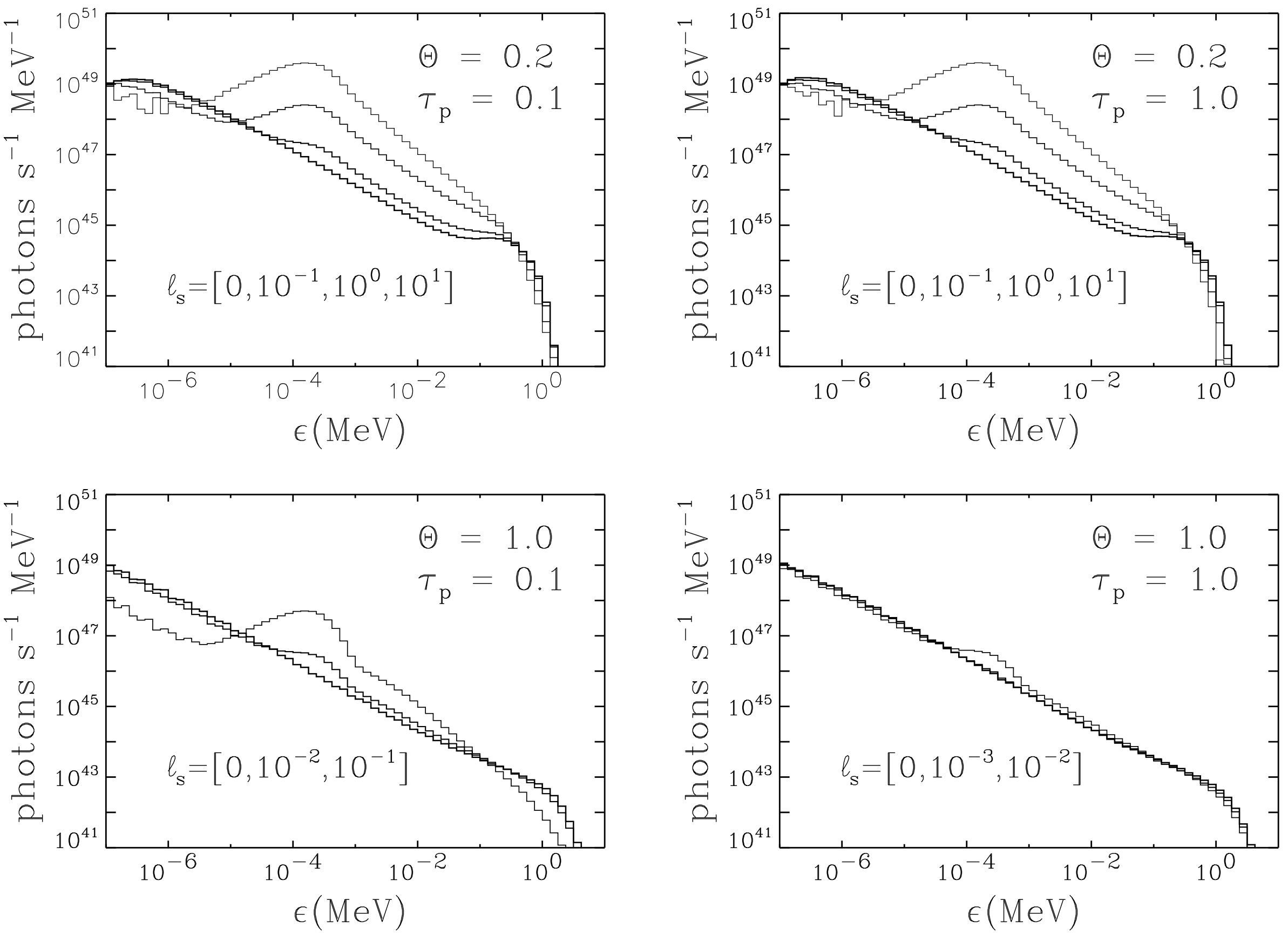

Fig. 7.— Spectra for the high-$z$ pair-balanced states of a plasma for various values of the soft photon injection taken to be a diluted blackbody with $kT = 0.1$ keV and compactness as indicated in the individual panels. The parameters $\Theta$ and $\tau_p$ are the same as those used in Fig. 5.

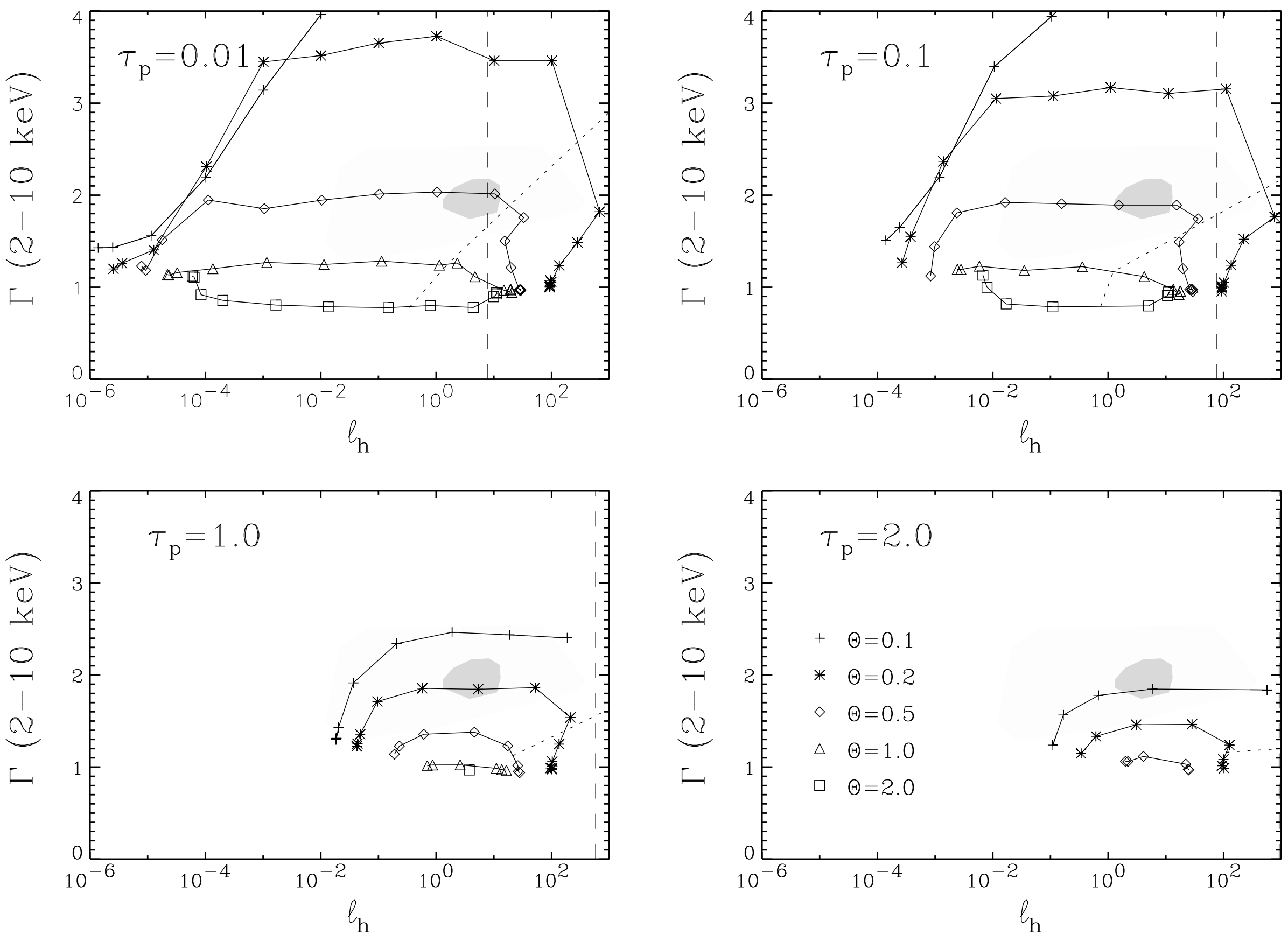

Fig. 8.— Photon 2-10 keV spectral index Γ as a function of emergent hard compactness $\ell_h$ for thermal plasmas in pair balance. The panels are labeled by the values of proton optical depth $\tau_p$. The curves represent isotherms with dimensionless temperature Θ given by the symbols in the legend. The light shaded region represents the envelope of the entirety of the AGN data in listed in Table 1 and the smaller dark region represents the Seyfert 1 "island" discussed in the text. The dotted curve delineates the boundary between the pair dominated ($z > 1$) and pair dilute ($z < 1$) regions, with the pair dominated solutions lying to the left. The vertical dashed line represents the efficiency limit of equation (6.4).

## 6. Comparison with Data of AGNs

We consider the sample of 21 AGNs listed in Table 1, which includes Seyferts, Narrow Emission Line Galaxies, and one Broad Line Radio Galaxy. Except for NGC 4151, which we classify separately because of its Sy 1.5 to Sy 2 transitions, all 2-10 keV luminosities and spectral indices are from the analysis of Ginga data by Nandra & Pounds (1994). They consider the effects of photoelectric absorption and reflection by cold matter near the source on the intrinsic source spectra. For NGC 4151, we use the luminosity derived by Awaki et al. (1991), who subtracts effects of photoelectric absorption by surrounding cold gas on the intrinsic spectrum of NGC 4151. Reflection is not important for NGC 4151, as shown by Maisack (1993) and Maisack & Yaqoob (1991). We use the measured X-ray variability doubling time scales $\Delta t$ quoted by Lightman & Zdziarksi (1987) and Done & Fabian (1989) to determine the hard compactnesses $\ell_h$ through the relation

$$\ell_h = (\frac{\sigma_T}{m_e c^4}) \left(\frac{E_u^{2-\Gamma} - E_l^{2-\Gamma}}{10^{2-\Gamma} - 2^{2-\Gamma}}\right) \left(\frac{L}{\Delta t}\right)_{2-10\mathrm{keV}} . \tag{6-1}$$

The middle term in parentheses is a bolometric correction to the 2-10 keV compactness, and depends on the spectral index and cutoff energies $E_l$ and $E_u$.

Figure 9 shows the location of the 21 AGNs in the $\Gamma - \ell_h$ plane. As can be seen, the intrinsic spectral indices cluster in the range $1.5 \lesssim \Gamma \lesssim 2.5$, and the hard compactnesses are found mostly in the range $0.01 \lesssim \ell_h \lesssim 10^2$. The value of $\ell_h$ may be a lower bound if the measured $\Delta t$ values are limited by statistics or incomplete sampling. The error bars indicate the uncertainty in the range of $\ell_h$ due to uncertainty in the values of $E_l$ and $E_u$. The lower limit corresponds to $E_l = 1$ keV and $E_u = 50$ keV, whereas the upper limit corresponds to to $E_l = 10$ eV and $E_u = 500$ keV. Note that there is a marked clustering of Seyfert 1 AGNs for $2 \lesssim \ell_h \lesssim 10$ and $1.8 \lesssim \Gamma \lesssim 2.2$. We call this region the "Seyfert 1 island," but more observations must be made to assess the significance of this clustering. (For a discussion of X-ray observations and variability time scales, see Mushotzky et al. 1993 and reference therein.) We also see that the NELGs and NGC 4151 evidently have smaller values of $\ell_h$ than the Seyfert 1s, although this may be a selection effect caused by the larger photoelectric absorption and weaker X-ray luminosity for these sources, which make the determination of $\Delta t$ more difficult.

In Figure 8, we plot shaded circles representing the approximate location of the Seyfert 1 island on the theoretical curves which give $\Gamma$ as a function of $\ell_h$ for thermal plasmas in pair balance. We suggest that thermal pair-balanced solutions for Seyfert AGNs are unlikely when $\tau_p \lesssim 0.01$ for two reasons. The first is due to the allowed limits on compactness. Clearly not all compactnesses are physically possible because, following Lightman & Zdziarski (1987),

$$\ell = 2\pi \left(\frac{L}{L_{Edd}}\right) \left(\frac{m_p}{m_e}\right) \left(\frac{R_S}{R}\right) \lesssim 10^3. \tag{6-2}$$

The final inequality holds if the source radiates at the Eddington limit and dissipates this power within 10 Schwarzschild radii. Actually the allowed compactnesses can be restricted even further, using the efficiency limits of Fabian (1979). Because the radiation escapes from a spherical source on the timescale $\Delta t \cong R(1+\tau_p/3)/c$, the energy liberated from the matter is bounded above by the expression

$$L\Delta t \lesssim \frac{4}{3}\pi R^3 n_p \eta m_p c^2, \tag{6-3}$$

where $\eta \equiv 10^{-1}\eta_{-1}$ is the efficiency of converting the rest mass of matter into radiant energy. Using the definition of $\ell$ and $\tau_p = n_p\sigma_T R$ in equation (6-3), one finds that

$$\ell \leq \frac{4}{3}\pi\eta\left(\frac{m_p}{m_e}\right)\frac{\tau_p}{1+\tau_p/3} \cong 770\eta_{-1}\frac{\tau_p}{1+\tau_p/3}. \tag{6-4}$$

Thus we see that for 10% efficiency, there are no allowed solutions for several high compactness AGNs if the thermal plasmas have $\tau_p \lesssim 0.01$. The 10% efficiency limits are shown in Figure 8 by the vertical dashed lines. In this argument we ignored the contribution of pairs to the scattering opacity in estimating the photon diffusion time $\Delta t \cong (1+\tau_p/3)R/c$ because pair densities can attain equilibrium on even faster time scales. We point out, however, that the arguments leading to equation (6-4) can be weakened by considering other geometries or by extracting energy from sources other than the rest mass of the accreting matter (Rees 1982).

The second reason that the low optical depth solutions are unlikely is due to the observed spectral smoothness and relative uniformity of measured spectral indices of Seyfert AGNs, compared to results from spectra calculated with small optical depths. When $\tau_p \lesssim 0.01$, the allowed solutions in the vicinity of the Seyfert 1 island show distinct spectral structure arising from individual scattering profiles, as shown in Figure 3 (see Pozdnyakov et al. 1983; Zdziarski et al. 1994). The large variations in spectral index expected in the low optical depth regime have not been seen in broadband X-ray measurements of Seyfert AGNs (e.g., Rothschild et al. 1983).

When $\tau_p \gtrsim 0.1$, the values of $\Gamma$ become increasingly uniform for a given temperature. We see from the location of the Seyfert 1 island that temperatures $\Theta \sim 0.4-0.6$ are consistent with the data, although the efficiency constraint still requires larger values of $\tau_p$ in some cases. At still larger values of $\tau_p$, smaller temperatures are required to give agreement between the observations and calculations. Thus we see that if Seyfert AGNs are modelled by thermal plasmas in pair balance, then temperatures $\Theta \gtrsim 0.6$, or $kT \gtrsim 300$ keV, are inconsistent with the data. Moreover, these solutions are invariably in the low-pair regime. This is made clear by noting that the AGNs lie to the left of the dotted curve indicating $z = 1$ in Figure 8, which separates the pair-dilute and pair-dominated regimes. Thus we predict that if Seyfert AGNs are thermal, then they should show spectral cutoffs corresponding to temperatures $\lesssim 300$ keV. We also conclude that observations of Seyfert AGNs strongly suggest that the plasmas responsible for the high energy emission are unlikely to be pair-dominated. The condition of pair dominance was used to reduce the number of parameters and establish a one-to-one correspondence between the observables and the physical parameters of the source (Ghisellini & Haardt 1994). But our comparison with the inferred compactnesses of AGNs shows that AGN data are found in the pair-dilute regime.

Present observations are not sufficiently sensitive to determine the energy of spectral cutoffs of

Table 1. Seyfert Data

| Name | Type | $\Gamma_x$ | $\log L_x$ | $\log \Delta t$ | $\ell_h$ |
|---|---|---|---|---|---|
| Mrk 335 | Sy 1 | 2.39±0.15 | 43.5 | 3.8 | 64±49 |
| Fairall-9 | Sy 1 | 1.94±0.09 | 44.5 | 5.5 | 4.8±2.1 |
| NGC 3227 | Sy 1 | 1.95±0.07 | 42.4 | 4.3 | 2.3±1.0 |
| Akn 120 | Sy 1 | 1.93±0.11 | 44.2 | 3.4 | 270±120 |
| NGC 3516 | Sy 1 | 2.08±0.09 | 43.0 | 4.0 | 4.1±2.1 |
| NGC 3783 | Sy 1 | 2.11±0.07 | 43.6 | 4.3 | 8.2±4.4 |
| NGC 4051 | Sy 1 | 1.84±0.09 | 41.5 | 3.0 | 4.6±2.0 |
| NGC 4593 | Sy 1 | 1.81±0.07 | 43.0 | 3.4 | 3.8±1.6 |
| MCG-6-30-15 | Sy 1 | 2.12±0.04 | 43.0 | 3.3 | 150±82 |
| NGC 5548 | Sy 1 | 1.81±0.02 | 43.6 | 4.3 | 7.8±3.4 |
| Mkn 841 | Sy 1 | 2.30±0.02 | 43.7 | 5.0 | 3.1±2.2 |
| Mkn 509 | Sy 1 | 1.86±0.02 | 44.4 | 5.7 | 2.3±1.0 |
| NGC 7469 | Sy 1 | 1.97±0.05 | 43.6 | 4.9 | 9.0±4.1 |
| NGC 4151 | Sy 1.5 | 1.48±0.04 | 42.7 | 4.6 | 1.2±0.7 |
| NGC 2110 | NELG | 1.93±0.11 | 43.0 | 4.9 | 0.47±0.21 |
| NGC 2992 | NELG | 2.30±0.21 | 42.7 | 5.1 | 0.067±0.046 |
| NGC 526A | NELG | 1.49±0.21 | 43.2 | 4.9 | 0.027±0.015 |
| NGC 7314 | NELG | 1.91±0.11 | 42.6 | 3.3 | 9.1±4.0 |
| NGC 5506 | NELG | 2.08±0.05 | 43.2 | 5.2 | 65±33 |
| MCG-5-23-16 | NELG | 1.86±0.12 | 43.1 | 5.0 | 0.24±0.10 |
| 3C 382 | BLRG | 1.73±0.11 | 44.3 | 5.4 | 3.4±1.6 |

$L_x$ is the 2-10 keV luminosity assuming $H_0 = 50$ km s$^{-1}$ Mpc$^{-1}$ and $\Gamma_x$ is the 2-10 keV underlying power law.

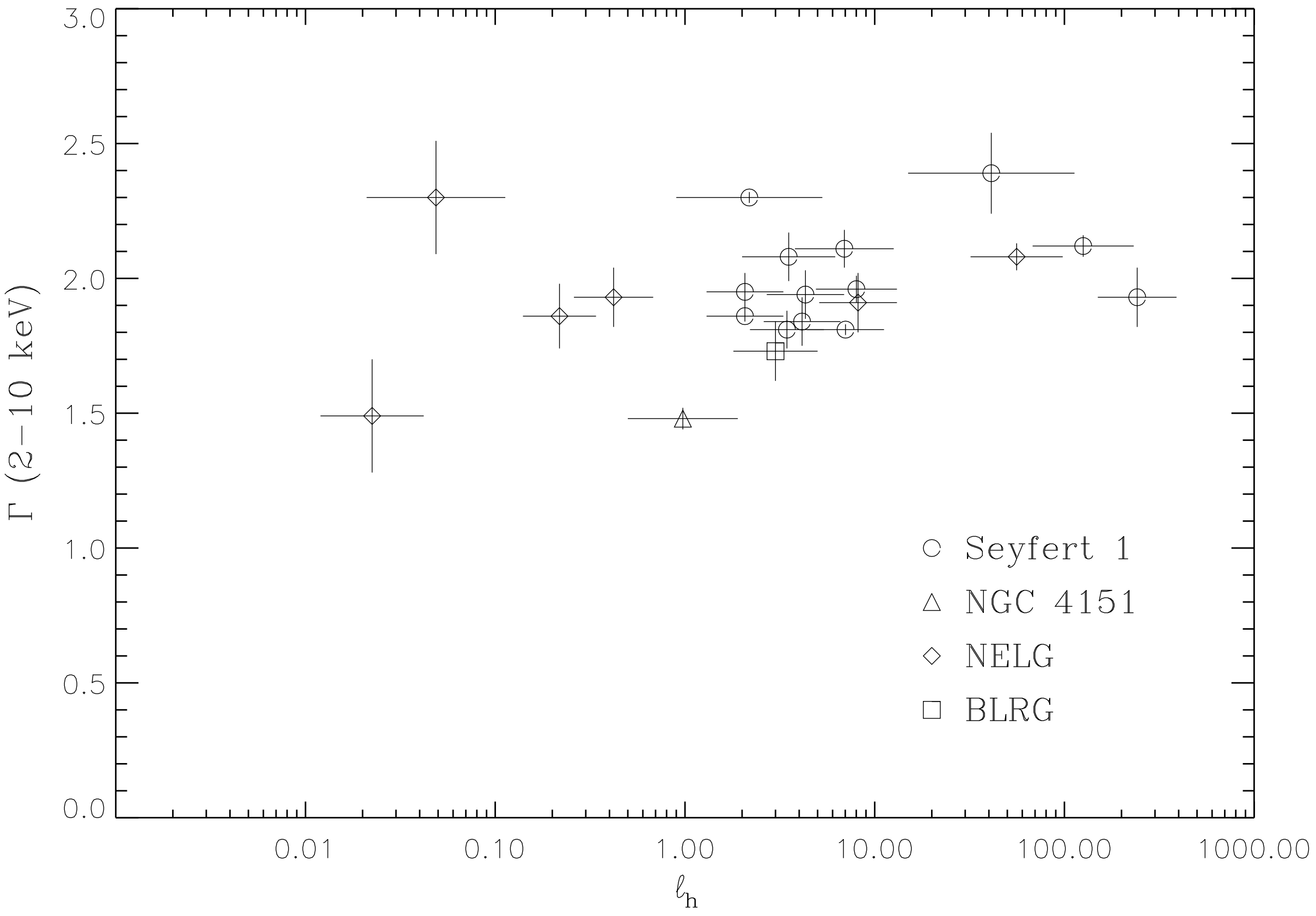

Fig. 9.— A plot of the photon spectral index versus hard compactness for the 21 AGN listed in Table 1.

Seyfert AGNS except in a few cases. OSSE observations of NGC 4151 (Maisack et al. 1993) show a strong spectral cutoff at $\sim$ 50 keV which is consistent with our prediction. NGC 4151, however, is not a typical Seyfert 1. IC 4329A is a standard Seyfert 1, but the presence of a reflection component makes it possible to fit the OSSE obersations (Madejski et al. 1994) with spectral cutoffs ranging from 300-700 keV. The average spectrum determined from OSSE observations of Seyfert 1 AGNs (Johnson et al. 1994) displays a cutoff at $\sim$ 50 keV, but it is important to take into account the reflection component found in such sources (Nandra & Pounds 1994). When this is done, analysis of OSSE observations of four radio quiet AGNs, excluding IC4329A, shows a spectral softening with average e-folding energy of several hundred keV (Zdziarski et al. 1995).

## 7. Summary

We have developed a Monte Carlo thermal Comptonization code which contains accurate $ep$, $ee$, and $e^+e^-$ bremsstrahlung and pair annihilation spectra. Pair production through $\gamma\gamma$ attenuation is calculated self consistently from the Comptonized angle and energy-dependent photon distribution. We also considered internally generated soft photons. We have compared our results for the case of pair free plasmas with the analytic expressions for thermal Comptonization found in the literature. We constructed a grid of solutions for pair-balanced plasmas and presented the results in the spectral index/compactness plane. We then compared our solutions with the X-ray data from a sample of 21 Seyfert AGNs and found that pair-dominated solutions are inconsistent with the data if the X-ray variability time scale gives an accurate measure of the compactness. Furthermore, we predict that the temperatures of central engines of Seyfert galaxies are $\lesssim$ 300 keV. This prediction can be tested with more sensitive gamma-ray observations of Seyfert galaxies.

We thank Dr. A. A. Zdziarski for useful discussions.

## A. Thermal Bremsstrahlung Gaunt Factor

We generate fitting functions accurate to within 8% for the Gaunt factors $g_{12}(x;\Theta)$ of ep, ee, and $e^+e^-$ thermal bremsstrahlung using results of numerical integrations of the emissivities given by Dermer (1986). The fitting functions reduce to asymptotic forms at nonrelativistic temperatures, extreme relativistic temperatures, and in the soft-photon regime. In the transrelativistic regime, Stepney & Guilbert (1983) and Haug (1987) also provide tables for ee and $e^+e^-$ thermal bremsstrahlung emissivities, respectively. The form of the Gaunt factors in the soft photon regime, $x \leq \Theta$, approaches the soft-photon Gaunt factors given by Svensson (1984):

$$g_{ep}^{sv}(x;\Theta) \;=\; \left(\frac{3\Theta}{2\pi}\right)^{1/2} \frac{\ln\left[4\eta(1+c_1\Theta)/u\right]}{e^{1/\Theta}K_2(1/\Theta)} \left(1+2\Theta+2\Theta^2\right) e^{u}, \qquad \text{(A1a)}$$

$$g_{ee}^{sv}(x;\Theta) = \left(\frac{3\Theta}{2\pi}\right)^{1/2} \frac{\ln\left[4\eta(c_2+c_3\Theta^2)/u\right]}{e^{1/\Theta}K_2(1/\Theta)} \left(\frac{3\cdot 2^{1/2}}{5}\Theta + 2\Theta^2\right) e^u, \tag{A1b}$$

$$g_{+-}^{sv}(x;\Theta) = \left(\frac{3\Theta}{2\pi}\right)^{1/2} \frac{\ln\left[4\eta(1+c_3\Theta^2)/u\right]}{e^{1/\Theta}K_2(1/\Theta)} 2\left(2^{1/2} + 2\Theta + 2\Theta^2\right) e^u, \tag{A1c}$$

where $u = x/\Theta$, $\eta = 0.56146$, $c_1 = 3.42$, $c_2 = 11.2$ and $c_3 = 10.4$. Corrections to these expressions are made using the functional form

$$g_{12}(x;\Theta) = g_{12}^{sv}(x;\Theta)/[1 + a_{12}(\Theta) + b_{12}(\Theta)u]. \tag{A2}$$

The coefficients $a_{12}(\Theta)$ and $b_{12}(\Theta)$ are given by

$$a_{ep}(\Theta) = \begin{cases} 0.13\Theta^{0.64} & \Theta \le 0.3 \\ 0.060 & 0.3 < \Theta \le 0.8 \\ 0.050\Theta^{-0.92} & \Theta > 0.8 \end{cases} \tag{A3a}$$

$$b_{ep}(\Theta) = \begin{cases} 1.60\Theta^{0.31} & \Theta \le 0.25 \\ 1.04 & 0.25 < \Theta \le 0.5 \\ 0.88\Theta^{-0.24} & \Theta > 0.5 \end{cases} \tag{A3b}$$

$$a_{ee}(\Theta) = \begin{cases} -0.24\Theta^{0.40} & \Theta \le 0.6 \\ -0.21 & 0.6 < \Theta \le 1.5 \\ -0.24\Theta^{-0.59} & \Theta > 1.5 \end{cases} \tag{A3c}$$

$$b_{ee}(\Theta) = \begin{cases} 0.53\Theta^{-0.27} & \Theta \le 1.5 \\ 0.46 & \Theta > 1.5 \end{cases} \tag{A3d}$$

$$a_{+-}(\Theta) = \begin{cases} 0.12\Theta^{0.51} & \Theta \le 0.5 \\ 0.082 & 0.5 < \Theta \le 1.5 \\ 0.12\Theta^{-1.1} & \Theta > 1.5 \end{cases} \tag{A3e}$$

$$b_{+-}(\Theta) = \begin{cases} 0.96\Theta^{0.12} & \Theta \le 0.5 \\ 0.68\Theta^{-0.31} & 0.5 < \Theta \le 1.5 \\ 0.60\Theta^{-0.095} & \Theta > 1.5. \end{cases} \tag{A3f}$$

We also give corrections for thermal bremsstrahlung in the hard photon regime, $\Theta < x \le 10\Theta$. For *ep* thermal bremsstrahlung with $\Theta \le 0.7$ we write the Gaunt factor in the form

$$g_{ep}(x;\Theta) = g_{ep}^{nr}(x;\Theta)[1 + a(\Theta) + b(\Theta)u + c(\Theta)u^2], \tag{A4}$$

where the nonrelativistic Gaunt factor is

$$g_{ep}^{nr}(x;\Theta) = \frac{3^{1/2}}{\pi}\exp(x/2\Theta)K_0(x/2\Theta), \tag{A5}$$

(Green 1959) and the coefficients $a(\Theta)$, $b(\Theta)$ and $c(\Theta)$ are given by

$$a(\Theta) = 1.6\Theta^{1.4} \tag{A6a}$$

$$b(\Theta) = \begin{cases} 0.36\Theta^{0.92} & \Theta \leq 0.4 \\ 0.14 & 0.4 < \Theta \leq 0.7 \end{cases} \quad \text{(A6b)}$$

$$c(\Theta) = 0.53\Theta^{1.97}. \quad \text{(A6c)}$$

For $\Theta > 0.7$ we have

$$g_{ep}(x;\Theta) = 1 + a(\Theta) + b(\Theta)u + c(\Theta)u^2, \quad \text{(A7)}$$

where

$$a(\Theta) = 2.3\Theta^{0.71} \quad \text{(A8a)}$$

$$b(\Theta) = \begin{cases} 0.22\Theta^{3.1} & 0.7 < \Theta \leq 1.0 \\ 0.22\Theta^{1.8} & 1.0 < \Theta \leq 2.0 \\ 0.35\Theta^{0.98} & \Theta > 2.0 \end{cases} \quad \text{(A8b)}$$

$$c(\Theta) = \begin{cases} 0.099\Theta^{1.4} & 0.7 < \Theta \leq 2.5 \\ 0.14\Theta^{1.1} & \Theta > 2.5. \end{cases} \quad \text{(A8c)}$$

For $ee$ thermal bremsstrahlung

$$g_{ee}(x;\Theta) = 1 + a(\Theta) + b(\Theta)u + c(\Theta)u^2, \quad \text{(A9)}$$

where

$$a(\Theta) = \begin{cases} 3.0\Theta^{1.15} & \Theta \leq 1.5 \\ 3.38\Theta^{0.814} & \Theta > 1.5 \end{cases} \quad \text{(A10a)}$$

$$b(\Theta) = \begin{cases} 0.41\Theta^{1.6} & \Theta \leq 0.5 \\ 0.28\Theta^{1.17} & \Theta > 0.5 \end{cases} \quad \text{(A10b)}$$

$$c(\Theta) = \begin{cases} 0.46\Theta^{2.4} & \Theta \leq 0.6 \\ 0.34\Theta^{1.77} & 0.6 < \Theta \leq 1.5 \\ 0.45\Theta^{1.06} & \Theta > 5.5 \end{cases} \quad \text{(A10c)}$$

In the case of $e^+e^-$ thermal bremsstrahlung with $\Theta \leq 0.45$ we have

$$g_{+-}(x;\Theta) = \frac{1}{0.9194} g_{+-}^{nr}(x;\Theta)[1 + a(\Theta) + b(\Theta)u + c(\Theta)u^2], \quad \text{(A11)}$$

with $g_{+-}^{nr}(x;\Theta) = 2\sqrt{2} g_{ep}^{nr}(x;\Theta)$ (Haug 1985) and

$$a(\Theta) = \begin{cases} 10.01\Theta^{2.49} & \Theta \leq 0.15 \\ 4.7\Theta^{2.09} & 0.15 < \Theta \leq 0.3 \\ 2.03\Theta^{1.36} & 0.3 < \Theta \leq 0.45 \end{cases} \quad \text{(A12a)}$$

$$b(\Theta) = -0.224\Theta^{0.63} \quad \text{(A12b)}$$

$$c(\Theta) = 1.29\Theta^{1.8} \quad \text{(A12c)}$$

And finally, for $\Theta > 0.45$ we have

$$g_{+-}(x;\Theta) = 1 + a(\Theta) + b(\Theta)u + c(\Theta)u^2, \quad \text{(A13)}$$

where

$$a(\Theta) = \begin{cases} 7.3\Theta^{0.88} & 0.45 < \Theta \leq 1.5 \\ 7.6\Theta^{0.76} & \Theta > 1.5 \end{cases} \quad \text{(A14a)}$$

$$b(\Theta) = 0.41\Theta^{1.34} \quad \text{(A14b)}$$

$$c(\Theta) = \begin{cases} 0.74\Theta^{1.55} & 0.45 < \Theta \leq 2.0 \\ 1.08\Theta^{0.98} & \Theta > 2.0 \end{cases}. \quad \text{(A14c)}$$